\newcommand{\hst}{\textit{HST}}
\newcommand{\spitzer}{\textit{Spitzer}}

\newcommand{\ks}{\hbox{$K_s$}}

\newcommand{\ha}{\hbox{H$\alpha$}}
\newcommand{\lesssim}{\mathrel{\hbox{\rlap{\hbox{\lower4pt\hbox{$\sim$}}}\hbox{$<$}}}}

\newcommand{\lsim}{\lesssim}
\newcommand{\gtrsim}{\mathrel{\hbox{\rlap{\hbox{\lower4pt\hbox{$\sim$}}}\hbox{$>$}}}}

\newcommand{\gsim}{\gtrsim}

\newcommand{\mcal}{\hbox{$\mathcal{M}$}}

\newcommand{\etal}{et al.}

\newcommand{\msol}{\hbox{$\mathcal{M}_\odot$}}

\newcommand{\lsol}{\hbox{$L_\odot$}}

\newcommand{\infinity}{\hbox{$\infty$}}

\newcommand{\jmk}{\hbox{$J$$-$$K_s$}}

\newcommand{\ujy}{\hbox{$\mu$Jy}}
\newcommand{\lir}{\hbox{$L_{\mathrm{IR}}$}}
\newcommand{\mone}{\hbox{$[3.6\mu\mathrm{m}]$}}
\newcommand{\mtwo}{\hbox{$[4.5\mu\mathrm{m}]$}}

\newcommand{\micron}{\hbox{$\mu$m}}

\documentclass[11pt,twoside]{article}

\usepackage{asp2004}
\usepackage{epsf}
\usepackage{epsfig}
\usepackage{lscape}

\begin{document}
\title{Spitzer Observations of Red Galaxies at High Redshifts}   
\author{Casey Papovich, for the GOODS and MIPS GTO teams}   
\affil{Steward Observatory, 933 N. Cherry Ave., Tucson, AZ 85721}    

\begin{abstract} 
I discuss constraints on star--formation and AGN in massive, red
galaxies at $z$$\sim$1--3 using \spitzer\ observations at
3--24~\micron.   In particular I focus on a sample of distant red
galaxies (DRGs) with \jmk$>$2.3 in the southern Great Observatories
Origins Deep Survey (GOODS--S) field.   The DRGs have  typical stellar
masses $\mcal$$\gsim$$10^{11}$~\msol.  Interestingly, the majority
($\gsim$50\%) of these objects have 24~\micron\ flux densities
$\gsim$50~\ujy.   At these redshifts massive galaxies undergo intense
(and possibly frequent) IR--active phases, which is in constrast to
lower--redshift massive galaxies.   If the 24~\micron\ emission in
these $z$$\sim$1--3 galaxies is attributed to star formation, then it
implies star--formation rates (SFRs) in excess of $\simeq$100~\msol\
yr$^{-1}$.    These galaxies have specific SFRs equal to or exceeding
the  global average value at that epoch.  Thus, this is an active
period in their assembly.   Based on their X--ray luminosities and
near-IR colors, as many as 25\% of the massive galaxies at  $z$$\gsim$
1.5 host AGN, suggesting that the growth of supermassive  black holes
coincides with massive--galaxy assembly.
\end{abstract}



\vspace{-25pt}
\section{Introduction}

\noindent Most ($\sim$50\%) of the stellar mass in galaxies today formed during
the short time between $z$$\sim$3 and 1 (e.g., Dickinson et al.\ 2003,
Rudnick et al.\ 2003).   Much of this stellar mass density
resides in massive galaxies, which appear at epochs prior to
$z$$\sim$1--2 (see, e.g., McCarthy
2004, Renzini 2006).  It is still unclear when and where the stars in
these galaxies formed.  The fashionable scenario is that
galaxies ``downsize'', with massive galaxies forming most of their
stars in their current configuration at early cosmological times, with
less--massive galaxies continuing to form stars to the present
(e.g., Bauer et al.\ 2005, Juneau et al.\ 2005).   But it is unclear
when and where the stellar mass in galaxies forms.  For example, it
may be that stars form predominantly in low--mass galaxies at high
redshifts, which then merge over time to form large, massive galaxies
at more recent times (e.g., Kauffmann \& Charlot 1998, Cimatti et al.\
2002).

At $z$$\lsim$1 most massive galaxies exist on a fairly prominent red
sequence (e.g., Blanton et al.\ 2003, Bell et al.\ 2004), are largely
devoid of star formation, evolve passively, and contain up to half of
the stellar--mass density.
However, hierarchical models generally predict colors for massive
galaxies at $z$$\sim$0 that are too blue compared to observations
(e.g., Somerville et al.\ 2001, Dav\'e et al.\ 2005).  Some
recent models suppress star formation in massive galaxies at late
times by truncating it at some mass threshold, or by invoking strong
feedback from AGN (e.g., Granato et al.\ 2001, Dav\'e et al.\ 2005,
Croton et al.\ 2006, Hopkins et al.\ 2006).   The morphologies of the most optically
luminous (and the most massive) galaxies transforms from ``normal''
early--type systems at $z$$\sim$1 to irregular systems at
$z$$\sim$2--3 (e.g., Papovich et al.\ 2005).   To
understand the assembly of these objects we need to study the
properties of massive galaxies at these earlier epochs.

In these proceedings, I focus on \spitzer\ observations at
3--24~\micron\ of massive galaxies at $z$$\sim$1.5--3 in the southern
Great Observatories Origins Deep (GOODS--S) field.   Combined with
\hst/ACS and ground--based data, the \spitzer\ data  provide
constraints on the star--formation and AGN activity in massive
galaxies at these epochs.  I discuss the properties of the masses and
IR activity in these galaxies, and I comment on the implications for their
assembly and evolution.

\vspace{-8pt}
\section{Stellar Masses and Star Formation in High--$z$ Massive Galaxies}

\noindent GOODS is a multiwavelength survey of two 10\arcmin$\times$15\arcmin\
fields.  The GOODS datasets include (along with other things) \hst/ACS
and VLT/ISAAC imaging, \textit{Chandra} X--ray observations
(Giavalisco et al.\ 2004), and recent \spitzer\ imaging (see
M.~Dickinson's contribution to these proceedings).  I make use of
these data for the work described here, as well as data from
\spitzer/MIPS 24~\micron\ in this field from time allocated to
the MIPS GTOs (e.g., Papovich et al.\ 2004).

Here, I focus on so--called distant red galaxies (DRGs)
selected with \jmk$>$  2.3~mag (Saracco et al.\ 2001, Franx et al.\
2003).   This color criterion identifies galaxies at $z$$\sim$2--3.5
whose light is dominated by passively evolving stellar populations
older than $\sim$250~Myr (i.e., with a strong Balmer/  4000~\AA\ break
between the $J$ and \ks--bands), and also heavily reddened
star--forming galaxies at these redshifts (F\"orster--Schreiber et
al.\ 2004, Papovich et al.\ 2006).  For the GOODS--S data, the
\jmk$>$2.3~mag selection is approximately complete to stellar masses
$\geq$$10^{11}$~\msol\ for passively evolving galaxies  (see
Papovich et al.\ 2006).

More than 50\% of the DRGs have 24~\micron\ detections with
$f_\nu(24\micron)$$\geq$50~\ujy.   Daddi et al.\ (2005), Reddy et al.\
(2006), and Webb et al.\ (2006) also find similar
24~\micron--detection rates for massive galaxies at $z$$\sim$2..
Interestingly, this implies that the \textit{majority} of massive galaxies at
$z$$\sim$2 emit strongly in the thermal IR ---
\textit{they are either actively forming stars, supermassive blackholes, or
both at this epoch}.  The 24~\micron\ emission at $z$$\sim$1.5--3
probes the mid--IR ($\sim$5--10~\micron), which broadly correlates
with the total IR, $\lir$$\equiv$$L(8-1000\micron)$ (e.g., Chary \& Elbaz
2001).  Figure~1 shows the inferred
\lir\ for the DRGs using the Dale et al.\ (2002) models to convert the
observed mid--IR to total IR luminosity.   Although there is inherent
uncertainty in this conversion, the 24~\micron\ flux densities for the
$z$$\sim$1.5--3 DRGs yield $\lir$$\approx$$10^{11.5-13}$~\lsol. If
attributed to star--formation, then this corresponds to SFRs of
$\approx$100--1000~\msol\ yr$^{-1}$ (e.g., Kennicutt 1998).

\begin{figure}[t]
\plottwo{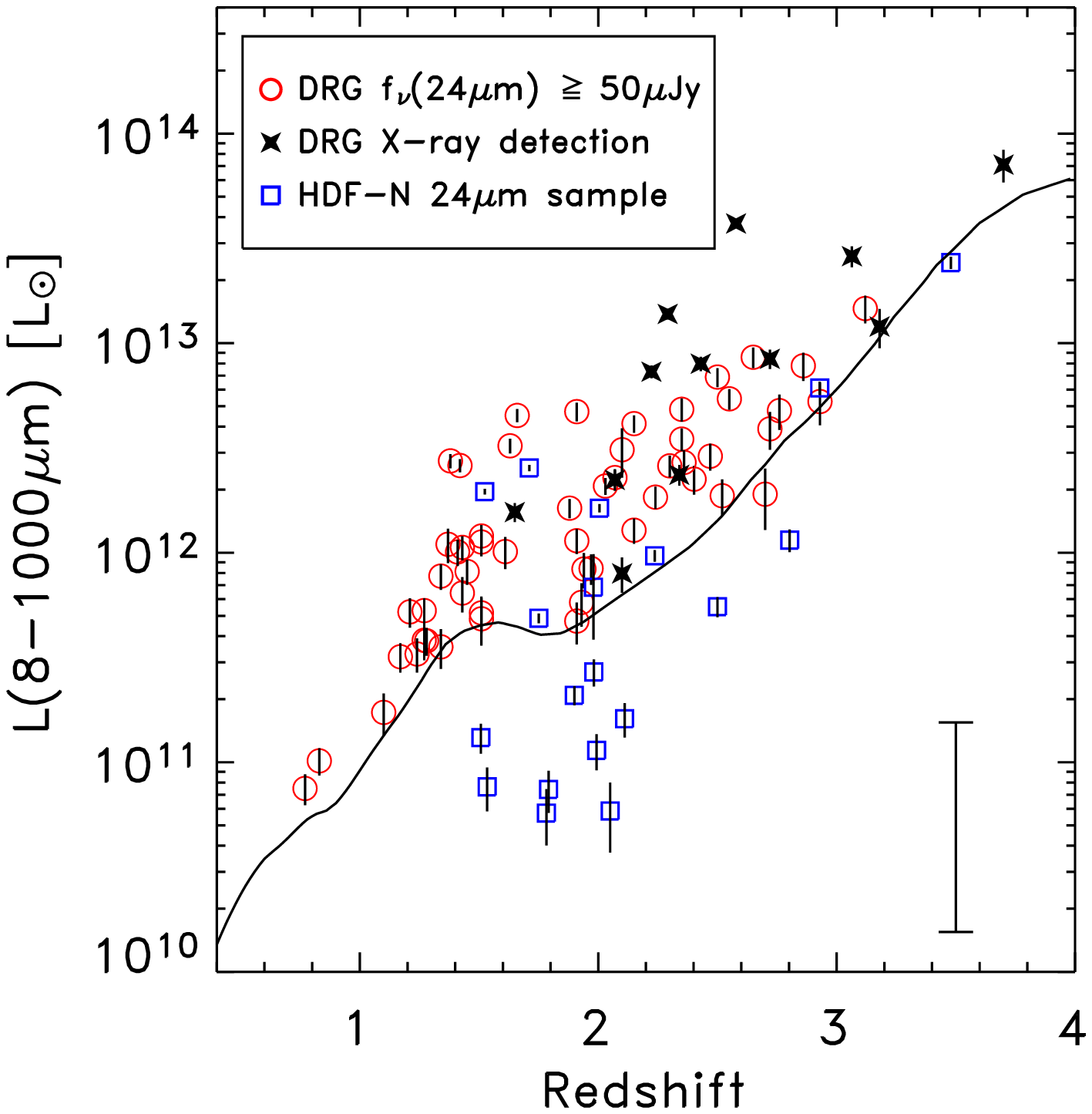}{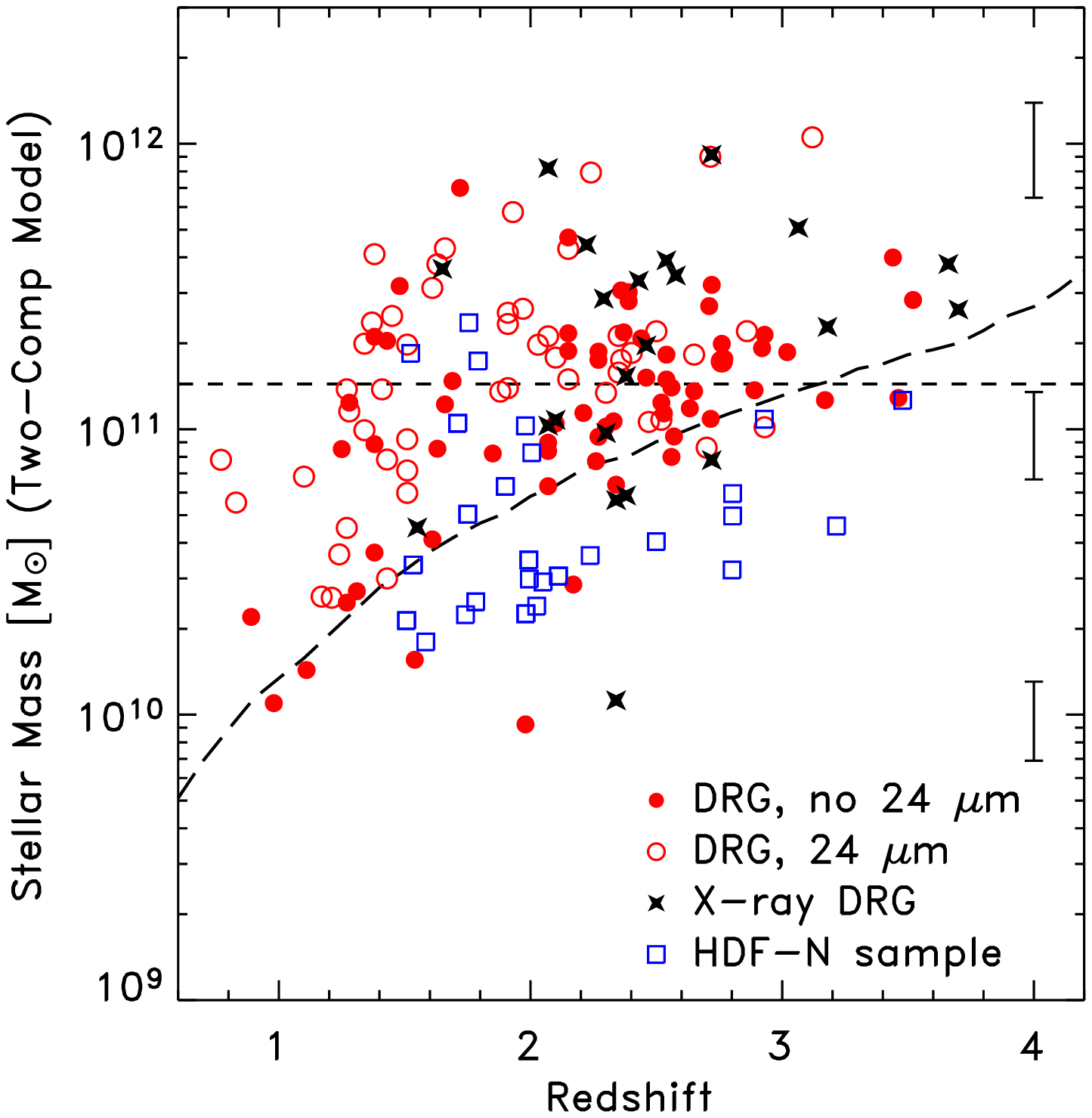}
   \caption{\textit{LEFT:} Total IR luminosities,
   $\lir$$\equiv$$L(8-1000)$, of galaxies inferred from their observed
   24~\micron\ emission.  Symbol definitions are inset.  The solid
   line denotes the 24~\micron\ 50\% completeness limit of the GTO
   data.   The estimated systematic error is $\approx$0.5~dex..
   \textit{RIGHT}: Stellar masses of galaxies inferred by fitting
   models to the galaxies' rest--frame UV--to--near-IR data.  The
   inset bars show the mean errors as a function of mass. The
   short--dashed line shows the characteristic present--day stellar
   mass (Cole et al.\ 2001); the long--dashed line shows the stellar
   mass limit for a passively evolving stellar population formed at
   $z$$\sim$$\infinity$ with $\ks$=23.2~mag. }
\vspace{-10pt}
\end{figure}

\begin{figure}[t]
\plottwo{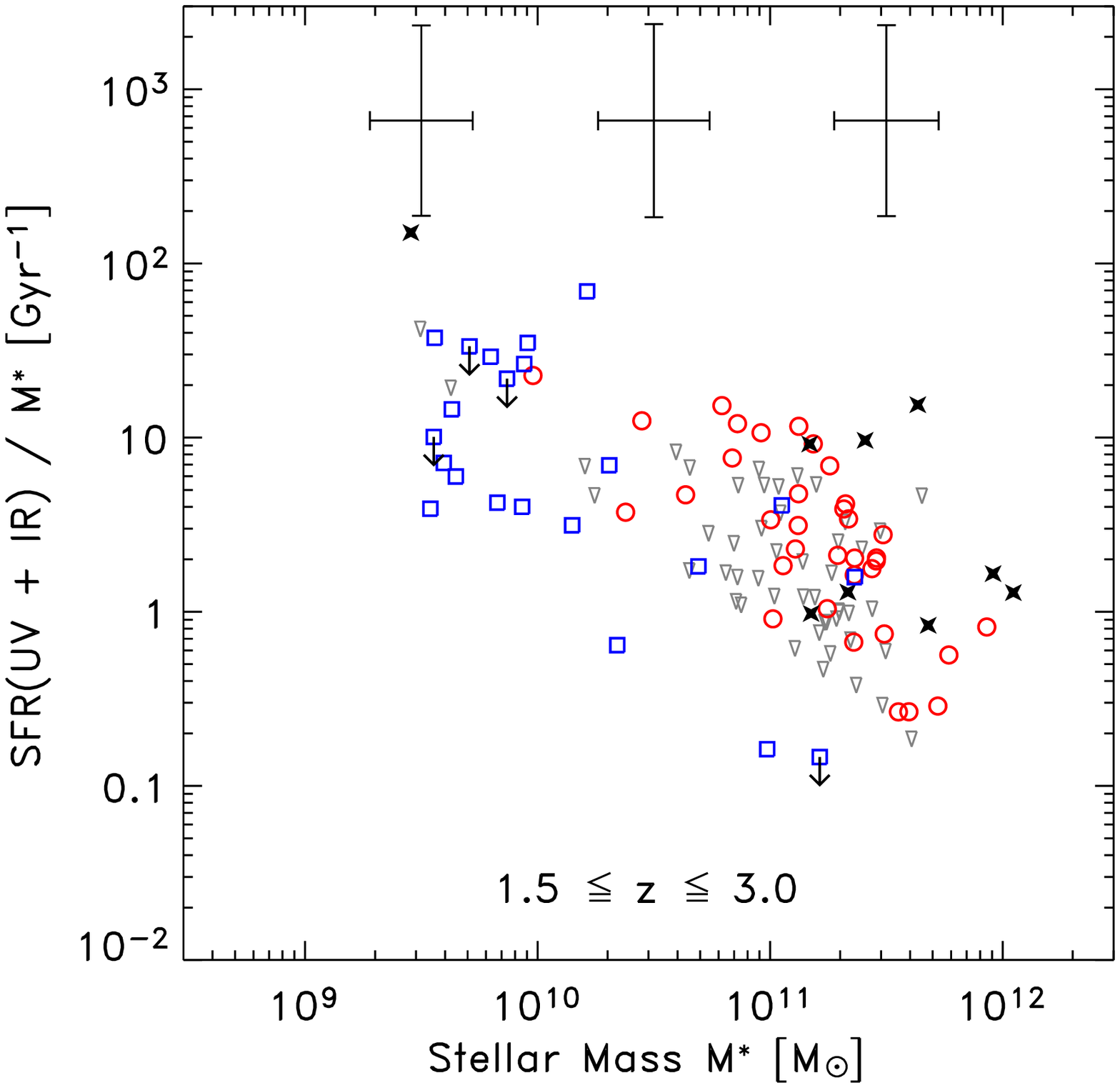}{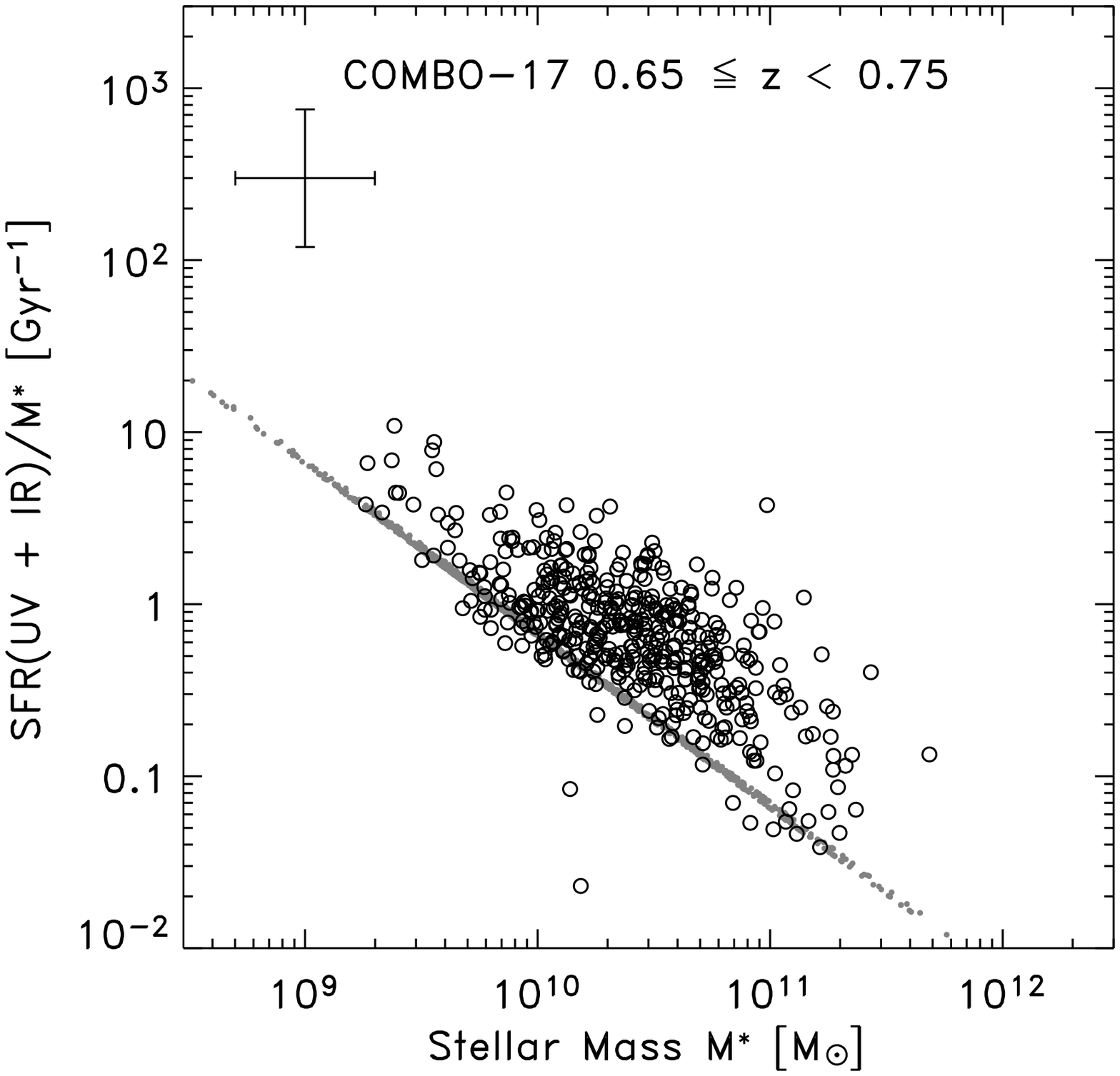}
   \caption{Specific SFRs as a function of galaxy stellar mass.   The
   Left panel shows the DRG and HDF--N galaxies with 1.5
   $\leq$$z$$\leq$ 3, with symbols the same as in figure~1.  The Right
   panel shows lower--redshift galaxies from COMBO--17.  Open circles
   show COMBO--17 galaxies with 24~\micron\ detections, small filled
   symbols show upper limits for galaxies undetected at 24~\micron.}
\vspace{-10pt}
 \end{figure}

Nearly all of the DRGs are detected in the deep \spitzer/IRAC data,
implying they have substantial stellar masses.  In Papovich et al.\
(2006), we modeled the DRGs by comparing their ACS,
ISAAC, and IRAC \mone\ \mtwo\  photometry to 
stellar--population synthesis models (Bru\-zu\-al \& Char\-lot 2003).
Although the modeling loosely constrains the ages, dust content, and
star--formation histories of the DRGs, it provides relatively robust
estimates of the galaxies' stellar masses, which are plotted in
figure~1 as a function of redshift.  Typical uncertainties for the
stellar masses for the full DRG sample are 0.1--0.3~dex.

Figure~2 shows the specific SFRs ($\Psi/\mcal$) derived from the
masses and SFRs for the DRGs, where the SFRs are derived from the
summed UV and IR emission.   The figure also shows the specific SFRs
for lower redshift galaxies  from COMBO--17 (Wolf et al.\ 2003), which
overlaps with the GTO 24~\micron\ data.  The massive galaxies with
$\mcal$$\geq$$10^{11}$~\msol\  at 1.5$\leq$$z$$\leq$3 have high specific SFRs, 
$\Psi/\mcal$$\sim$0.2--10 Gyr$^{-1}$ (excluding X--ray sources).   In
contrast, at $z$$\lsim$ 0.75 galaxies with
$\mcal$$\geq$$10^{11}$~\msol\ have much lower specific SFRs,
$\Psi/\mcal$$\sim$0.1--1 Gyr$^{-1}$.

In Papovich et al.\ (2006), we defined the integrated specific SFR as
the ratio of the sum of the SFRs, $\Psi_i$, to the sum of their
stellar masses, $\mcal_i$, $\Upsilon$$\equiv$$\sum_i
\Psi_i / {\sum_i\mcal_i}$,  summed over all $i$ galaxies.   Figure~3 shows the
integrated specific SFRs for DRGs at $z$$\sim$1.5--3.0 and COMBO--17
at $z\sim 0.4$ and 0.7, all with $\mcal \geq 10^{11}$~\msol. 
The error box indicates the affect of assumptions in the SFRs and AGN
activity in the DRGs (see further discussion below, and in Papovich et
al.\ 2006).   The integrated specific SFR in galaxies with
\mcal$>$$10^{11}$~\msol\ declines by more than an order of magnitude
from $z$$\sim$1.5--3 to $z$$\lsim$0.7.

 \begin{figure}[t]
  \centering
   \includegraphics[width=0.6667\linewidth]{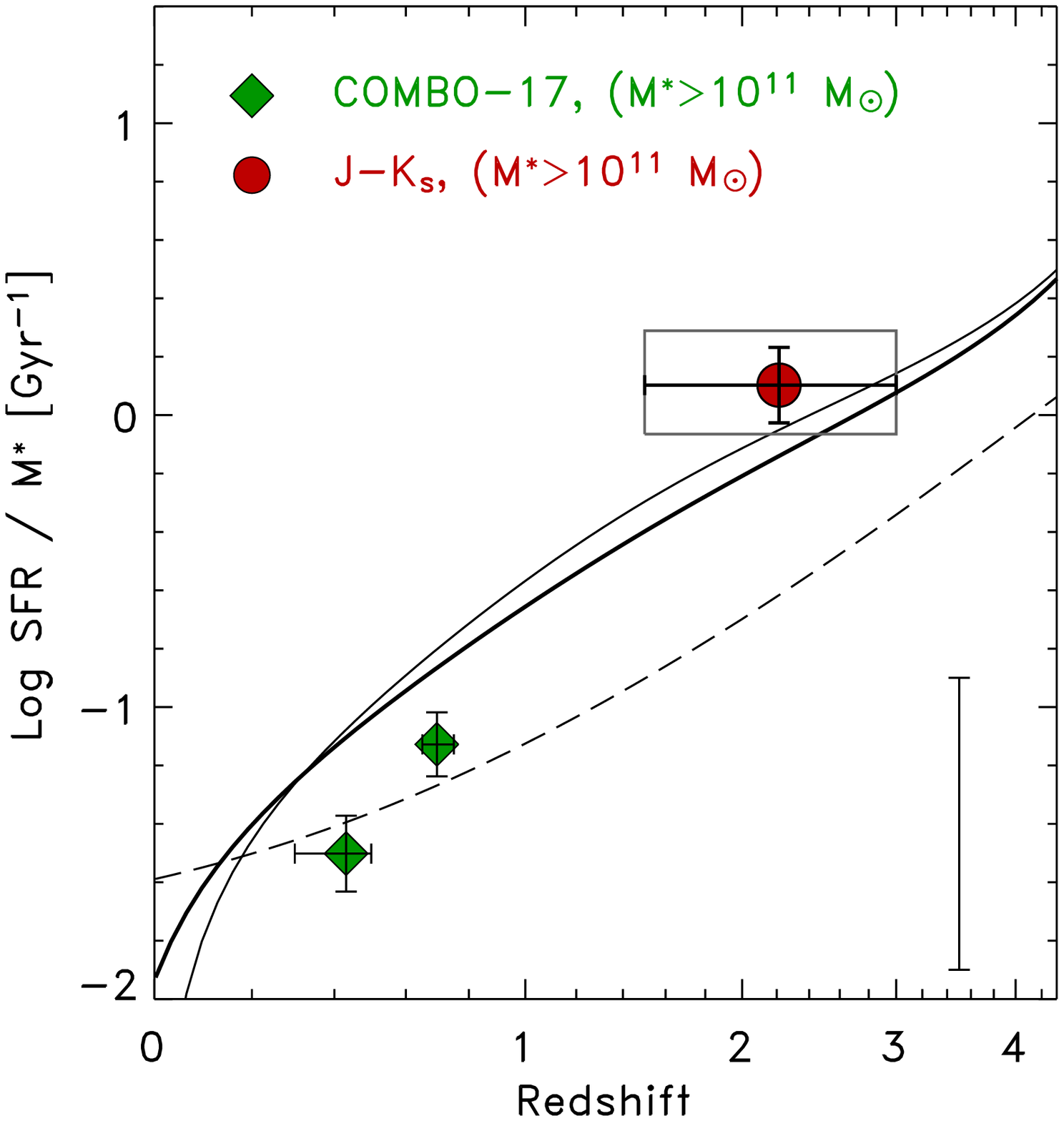}
 \caption{Evolution of the integrated specific SFR, the ratio of the
  total SFR to the total stellar mass (from Papovich et al.\ 2006).
  The curves show the expected evolution from the global SFR density
  (solid lines, Cole et al.\ 2001, thick line includes correction for
  dust extinction; dashed line, Hernquist \& Springel 2003).  Data
  points show results for galaxies with $\geq$$10^{11}$ \msol.
  Filled circle corresponds to the DRGs; filled diamonds correspond to
  the COMBO--17 galaxies. The inset error bar shows an estimate on the
  systematics.}
\vspace{-10pt}
   \end{figure}

The curves in figure~2 show the specific SFR integrated over all
galaxies (not just the most massive);  this is the ratio of the cosmic
SFR density to its integral, $\dot{\rho}_\ast /
\int \dot{\rho}_\ast\, dt$.   Although there is a decrease in the global
specific SFR with decreasing redshift, the  evolution in the
integrated specific SFR in massive galaxies is accelerated.   The
implication is that \textit{at $z$$\gsim$1.5, massive galaxies are
rapidly forming their stars, whereas by  $z$$\lsim$1.5 the specific
SFRs of massive galaxies drops rapidly, and lower--mass galaxies
dominate the cosmic SFR density} .

\vspace{-5pt}
\section{The Contribution of AGN to the mid--IR Emission}

\noindent Many ($\sim$15\%) of the massive galaxies at $z$$\sim$1.5--3
are detected in the deep X--ray data, and these objects tend to have
high inferred IR luminosities and specific SFRs (see figs.~1 and
2).  Indeed, the fraction of X--ray detected DRGs rises with
increasing luminosity ($>$50\% at $\lir$$\gsim$$10^{13}$~\lsol),
similar to the detection rate of sub--mm galaxies (Alexander et al.\
2005).   At these redshifts the X--ray fluxes imply the presence of an
AGN with $L_X$$\gsim$$10^{42}$~erg~s$^{-1}$.   Many authors are also
finding AGN candidates based on rest--frame near--IR colors, where the
AGN is presumably heavily obscured by gas and dust so that it is
missed in deep ($\gsim$1~Msec) X--ray surveys (e.g., Donley et al.\
2005, Stern et al.\ 2005, Alonso--Herrero et al.\ 2006, Barmby et al.\
2006).   Approximately 10\% of the X-ray \textit{un}detected DRGs have
ACS--through--IRAC colors consistent with dust--enshrouded AGN.
Combined with the 15\% of DRGs detected in the X--rays, up to 25\% of
the DRG population host AGN (see also Papovich et al.\ 2006).

If AGN contribute to the observed 24~\micron\ emission in galaxies at
$z$$\sim$1.5--3, then they can affect the inferred IR luminosities.
For example, using an IR template for Mrk~231 instead of a
star--forming galaxy with $\lir$$\gsim$$10^{13}$~\lsol\ would reduce
the IR luminosity for $z$$\sim$1.5--3 galaxies by a factor
of $\sim$2--5.   To limit this effect on the evolution of the
integrated specific SFRs, the error box in figure~3 shows how the
result changes if for galaxies with putative AGN we set the SFR to
zero.  The high AGN occurrence in DRGs suggests  that
massive galaxies at $z$$\sim$1.5--3 simultaneously form stars and grow
supermassive black holes.

However, even when AGN are present it is unclear whether
star--formation or the AGN activity domninates the bolometric IR
luminosity.  Although $\sim$80\% of sub--mm galaxies have X--ray
detections (Alexander et al.\ 2005), the IR to X--ray luminosity
ratios are up to an order of magnitude higher than what is expected
for AGN alone.  This suggests that both star--formation and AGN
contribute to the bolometric emission.  Similarly, in the near--IR
spectrum of a $z$$\sim$2.5 DRG, Frayer et al.\ (2003) find that the
galaxy nucleus has a low \ha\ to [N{\small II}] flux ratio consistent
with ionization from an AGN.   However, \ha\ is spatially resolved in
their spectrum and the [N{\small II}] line strength drops off in the
off--nucleus spectrum.   This implies extended star--formation beyond
the nucleus, which presumably contributes to the inferred IR
luminosity.  Both AGN and star--formation occur simultaneously in
high--redshift IR--detected galaxies and both probably contribute to
the IR emission.

\vspace{-5pt}
\section{Summary}

In this contribution, I discussed star--formation and AGN activity in
massive galaxies ($\gsim$$10^{11}$~\msol) at $z$$\sim$1--3 using
observations from \spitzer\ at 3--24~\micron.   The majority
($\gsim$50\%) of these objects have $f_\nu(24\micron)$$\geq$50~\ujy,
which if attributed to star formation implies SFRs of
$\gsim$100~\msol\ yr$^{-1}$.    Galaxies at $z$$\sim$1.5--3 with
$\mcal$$\geq$$10^{11}$~\msol\ have specific SFRs equal to or exceeding
the global average value.  In contrast, massive galaxies at
$z$$\sim$0.3--0.75 have specific SFRs less than the global average,
and more than 10$\times$ lower than that at $z$$\sim$1.5--3.   By
$z$$\lsim$1.5 massive galaxies have formed most of their stellar mass,
and lower--mass galaxies dominate the SFR density.   At the same time,
as many as 25\% of the massive galaxies at $z$$\gsim$1.5 host AGN.
The high  AGN occurrence at $z$$\sim$ 1.5--3 provides evidence that
massive galaxies are simultaneously forming stars and growing
supermassive black holes.   This may provide the impetus for the
present--day black-hole--bulge-mass relation and/or provide the
feedback necessary to squelch star--formation in such galaxies, moving
them onto the red sequence.

Lastly, while on average
high--redshift galaxies have IR spectral energy distributions
consistent with local templates, individually there remains significant
uncertainty.  Future work is needed to understand the
distribution between the mid--IR (rest--frame 5--15~\micron) emission
and total IR luminosity in $z$$\sim$1.5--3 galaxies.  
\textit{Herschel} Space Observatory (and eventually \textit{SAFIR})
will mitigate this problem by measuring the far--IR emission of
distant galaxies directly.

\acknowledgements 

I wish to thank the conference organizers for the invitation to
present this material, and for their work in planning a very
successful meeting. I am grateful for my colleagues on the MIPS GTO
and GOODS teams for their continued collaboration, in particular I am
indebted to L.~Moustakas, M.~Dickinson, E.~Le~Floc'h, E.~Daddi, and
G.~Rieke.   Support for this work was provided by NASA through the
Spitzer Space Telescope Fellowship Program, through a contract issued
by JPL/Caltech under a contract with NASA.


\end{document}